\documentclass[12pt, twocolumn]{openjournal}

\usepackage{lipsum}
\usepackage{xcolor}
\usepackage{textgreek}
\usepackage[utf8]{inputenc}
\usepackage[english]{babel}
\usepackage{hyperref}
\hypersetup{
    colorlinks=true,
    linkcolor=blue,
    filecolor=blue,      
    urlcolor=red,
    citecolor=blue,
}
\usepackage{color,colortbl}
\usepackage{tensind}
\tensordelimiter{?}
\DeclareGraphicsExtensions{.bmp,.png,.jpg,.pdf}
\usepackage{verbatim}
\usepackage[normalem]{ulem}
\usepackage{orcidlink}
\usepackage{soul}
\usepackage{bm}
\usepackage{amsmath}
\usepackage{float}
\usepackage{color}
\usepackage{xcolor}

\graphicspath{ {./figs/} }

\begin{document}
\title{The Jackknife method as a new approach to validate strong lens mass models}

\author{Shun Nishida$^{1}$\orcidlink{0009-0005-6923-5590}}
\author{Masamune~Oguri$^{1, 2}$\orcidlink{0000-0003-3484-399X}}
\author{Yoshinobu Fudamoto$^{2,3}$\orcidlink{0000-0001-7440-8832}}
\author{Ayari Kitamura$^{4}$}

\email{24wm2121@student.gs.chiba-u.jp}
\affiliation{$^{1}$Department of Physics, Graduate School of Science, Chiba University, 1-33 Yayoi-Cho, Inage-Ku, Chiba 263-8522, Japan}
\affiliation{$^{2}$Center for Frontier Science, Chiba University, 1-33 Yayoi-cho, Inage-ku, Chiba 263-8522, Japan}
\affiliation{$^{3}$Steward Observatory, University of Arizona, 933 N. Cherry Avenue, Tucson, AZ 85721, USA}
\affiliation{$^{4}$Astronomical Institute, Tohoku University, Aoba-ku, Sendai, Miyagi 980-8578, Japan}

\begin{abstract}
The accuracy of a mass model in the strong lensing analysis is crucial for unbiased predictions of physical quantities such as magnifications and time delays. While the mass model is optimized by changing parameters of the mass model to match predicted positions of multiple images with observations, positional uncertainties of multiple images often need to be boosted to take account of the complex structure of dark matter in lens objects, making the interpretation of the chi-square value difficult. We introduce the Jackknife method as a new method to validate strong lens mass models, specifically focusing on cluster-scale mass modeling. In this approach, we remove multiple images of a source from the fitting and optimize the mass model using multiple images of the remaining sources. We then calculate the multiple images of the removed source and quantitatively evaluate how well they match the observed positions. We find that the Jackknife method performs effectively in simulations using a simple model.  We also demonstrate our method with mass modeling of the galaxy cluster MACS J0647.7+7015. We discuss the potential of using the Jackknife method to validate the error estimation of the physical quantities by the Markov Chain Monte Carlo.
\end{abstract}


\maketitle

\section{Introduction}
\label{sec:intro}
The observation of galaxy cluster strong lensing is crucial for cosmology and for studying distant galaxies. By analyzing strong lensing observations, we can estimate the mass distribution of galaxy clusters.
The mass model derived from the analyses enables us to predict magnifications of lensed sources, which in turn allows us to infer intrinsic physical quantities of distant galaxies \citep[e.g.,][]{2012Natur.489..406Z, 2013ApJ...762...32C, 2023Sci...380..416W, 2024Natur.632..513A}, to predict the appearance timing of lensed supernovae \citep[e.g.,][]{2016ApJ...819L...8K,2021NatAs...5.1118R,2022Natur.611..256C}, and to constrain the Hubble constant from time delay measurements \citep[e.g.,][]{2023ApJ...959..134N,2023Sci...380.1322K, 2025ApJ...979...13P}. 
If the mass model is incorrect, 
we obtain biased estimates of these physical quantities.
This is why the accurate and precise mass modeling is essential in the strong lensing analysis. 

In the strong lensing analysis, we use observed data such as redshifts 
and positions of multiple images, and optimize a mass model by adjusting mass model parameters to reproduce the observed positions of the multiple images.
The fitting process minimizes the chi-square $\chi^2$ calculated from the difference between the observed and model-predicted multiple image positions.
If the model is valid, the minimized chi-square follows a $\chi^2$-distribution with the correct degrees of freedom (DoF) $\nu$, allowing the model validation by examining whether the reduced chi-square $\chi^2/\nu$ is approximately 1.

The positional error $\sigma$ in the denominator of $\chi^2$ is typically taken to be the measurement error. 
In the case of strong lensing analysis, however, reproducing multiple image positions within the measurement uncertainty is sometimes challenging because the complex dark matter substructure that is not included in mass modeling perturbs multiple image positions, which effectively act as an additional positional uncertainty. This positional error due to the complex dark matter substructure can be much larger than the measurement error, especially in the case of cluster strong lensing.
Therefore we need to 
include both a measurement error and the error caused by the complex dark matter substructure in the positional error.
However, an issue is that we have no clue to know the true value of the error caused by the complexity of the dark matter distribution because we cannot know the exact complexity of each dark matter halo of our interest.
This means that the value of chi-square has uncertainty 
and it is not appropriate to validate the mass model solely using the reduced chi-square $\chi^2/\nu$.

Until now, the mass model has been validated by various methods.
For instance, attempts to validate mass models have been made by comparing results from different lens model teams and software
\citep{2016ApJ...819L...8K, 2017MNRAS.465.1030P, 2020MNRAS.494.4771R}.
Another validation method is based on the analysis on mock strong lens data constructed with ray-tracing simulations \citep{2017MNRAS.472.3177M}, from which the accuracy and precision of the mass model for each software for strong lensing is tested and evaluated.
Additionally, the accuracy of mass models can be tested by comparing results with different sets of multiple images. \citet{2017MNRAS.470.1809A, 2019MNRAS.488.3251S} use mass models with and without newly observed multiple images and check the difference of the results. \cite{2018ApJ...863...60R} use the previous set of multiple images for optimizing the mass model, and predict the newly added set to check whether the model can accurately reproduce positions of the new multiple images.  \citet{2021MNRAS.508.5587R, 2025OJAp....8E..37P} split the sample into multiple bins based on certain criteria and conclude that lens models are neither converging nor diverging. \citet{2015ApJ...811...70R} measure the magnification factor of a lensed Type Ia supernova behind Abell 2744 and use it for testing the accuracy of various mass models. Various information criteria such as the Bayesian Information Criterion are sometimes used to compare different mass models \citep[e.g.,][]{2019A&A...632A..36C,2021A&A...645A.140B}.

However, these existing methods are not sufficient for a systematic and quantitative validation of mass models. The comparison across results of different lens modeling teams and software lacks clear validation criteria. Even though the results of the mass models from the different teams and software match, it does not guarantee the accuracy of the mass models.
The validation method with mock strong lens data is also 
not enough. The simulation for creating mock data can be quite different from actual galaxy clusters. 
Given the diversity of clusters, it also remains uncertain which simulated clusters should be matched with which strong lensing clusters in observations.
Furthermore, while the root mean square (RMS) of differences between observed and model-predicted multiple image positions is commonly used to quantify the precision of the mass model, we caution that the good RMS does not guarantee the good accuracy of the mass model, simply because the RMS quantifies the precision rather than the accuracy. For instance, some free-form methods \citep{2008ApJ...681..814C, 2023ApJ...951..140C} achieve a very small RMS but it does not necessarily mean that such models predict physical quantities such as magnifications and time delays very accurately.

Given the increasing importance of cluster strong lensing for studies of distant objects as well as for cosmology, we need a new method to validate strong lens mass models. In this paper, we propose the Jackknife method as a new validation method. 
We validate the mass model in terms of the following perspective: how accurately the mass model predict multiple image positions that are not used to construct the mass model. We test out new approach with the analysis on simple mock strong lens data. We also apply our method with an observed strong lens system to demonstrate the feasibility.

The outline of the paper is as follows.
In Sect.~\ref{sec:jackknife_method}, we explain the chi-square method in the strong lens analysis and introduce the Jackknife method as a new validation method.
In Sect.~\ref{sec:simulations}, we describe the set-up of the simulation for testing the effectiveness of the Jackknife method.
In Sect.~\ref{sec:demonstration}, we present the result of a demonstration using actual observations.
In Sect.~\ref{sec:discussion}, we investigate whether the Jackknife method has the potential to validate the error estimation of the physical quantities by the Markov Chain Monte Carlo (MCMC).
In Sect.~\ref{sec:conclusion}, we summarize our study and discuss future prospects. 

\section{The Jackknife method}
\label{sec:jackknife_method}
\subsection{The chi-square method}
In the strong lens analysis,
observed data such as redshifts and sky positions of multiple images are used to constrain and optimize mass models.
By assuming sources and lens models, and specifying parameters for model fitting, 
we optimize the mass model to match the observations with model predictions by adjusting fitting parameters. 

Suppose the fitting process involves $P$ observed multiple images $\bm{\theta}^{\rm obs}_i=(x_i, y_i)$ and a mass model with $Q$ parameters. We optimize the mass model using the chi-square defined by
\begin{equation}
    \chi^2 = \sum_{i=1}^{P}\frac{|\bm{\theta}_i^{\rm{obs}}-\bm{\theta}_i^{\rm{model}}(a_1,\cdots, a_Q)|^2}{\sigma_i^2},
    \label{eq:chi^2}
\end{equation}
where $\bm{\theta}_i^{\rm{model}}$ is the model-predicted sky locations of multiple images of the $i$-th image, $a_1,\cdots, a_Q$ denote mass model parameters, and $\sigma_i$ are the uncertainties of sky locations of multiple images.

Eq.~(\ref{eq:chi^2}) is the squared positional difference between the observed and model-predicted multiple images divided by the positional error $\sigma_i$, summed over all multiple images.
The model parameters are determined so that the chi-square is minimized. 
If the model and the error are valid, the chi-square follows $\chi^2$-distribution with degrees of freedom (DoF) $\nu = 2P - Q$ defined as
\begin{equation}
    f_{\nu}(x) = \frac{1}{2^{\frac{\nu}{2}}\Gamma(\frac{\nu}{2})}x^{\frac{\nu}{2}-1}e^{-\frac{x}{2}},
\end{equation}
where $\Gamma(x)$ is the Gamma function.
Since the $\chi^2$-distribution has a peak near degrees of freedom $\nu$,
one way to validate the model is to check whether the reduced chi-square $\chi^2/\nu$ is approximately 1 or not.
If the mass model is valid and the positional error is accurately estimated, $\chi^2$ follows the $\chi^2$-distribution with degrees of freedom $\nu$, and $\chi^2/\nu$ should be close to unity.

In the cluster strong lensing analysis, if we adopt the uncertainties in Eq.~(\ref{eq:chi^2}), $\sigma_i$, from positional uncertainties of actual measurements (hereafter, $\sigma_{\rm mes}$), the reduced chi-square of the best-fitting model tends to have values much larger than $1$, because of the additional positional uncertainty $\sigma_{\rm sub}$ caused by the complexity of the dark matter distribution in the lensing cluster (hereafter, $\sigma_{\rm{eff}}$).
Since $\sigma_{\rm sub}$ dominates over $\sigma_{\rm mes}$ in many cases, the effective total positional uncertainty $\sigma_{\rm eff}$ is written as
\begin{align}
    \sigma_{\rm{eff}}^2 &= \sigma_{\rm{sub}}^2 + \sigma_{\rm{mes}}^2 \approx \sigma_{\rm{sub}}^2.
\end{align}

Ray-tracing simulations in $N$-body simulations \citep[e.g.,][]{2017MNRAS.472.3177M} provide the estimation of a typical value for $\sigma_{\rm sub}$. However, galaxy clusters and their inherent dark matter distributions are so complex that there exists a large scatter in the complexity of the dark matter distribution \citep[e.g.,][]{2015PASJ...67...61I}.
Any mismatch between the assumed and actual $\sigma_{\rm sub}$ can result in an additional systematic uncertainty when deriving a mass model for each cluster.

We do not know the true value of $\sigma_{\rm{eff}}$ because we cannot know the exact structure of each clusters. 
Thus, the validity of mass models cannot be judged by the value of the reduced chi-square.
The uncertainty of the chi-square leads to the uncertainty of physical quantities, such as magnifications and time delays predicted by the best-fitting mass model. 

\subsection{The Jackknife method}
As a new validation criterion, we develop a new method to examine a predictive power of each mass model.
The predictive power means the ability of a model to accurately predict physical quantities such as positions of multiple images, magnifications, and time delays, which are not included as observational constraints when deriving the mass model.
Here we caution the difference between accuracy and precision. 
The accuracy refers to how close a measured value is to the true value. 
The precision refers to how close multiple measurements are to each other, regardless of whether they are close to the true value. 
We categorize models as correct and incorrect based on the accuracy i.e., the predictive power.
If the model has high predictive power, the model is considered correct.
If the model has low predictive power, the model is considered incorrect.

In this paper, we use the Jackknife method to quantify the predictive power of the mass model. This method evaluates how well a mass model can predict positions of multiple images that are excluded from the fit. The specific procedure of the Jackknife method is shown below.

We assume a situation where we find multiple images of $R$ sources from observations.
\renewcommand{\labelenumi}{(\roman{enumi})}
\begin{enumerate}
    \item We remove all multiple images that belong to one of $R$ sources in observations.
    \item We optimize the mass model using the remaining $R-1$ sources.
    \item We calculate sky positions of the multiple images of the source that is 
    removed from the fitting, using the best-fitting model that is derived in (ii).
    \item We compare the observed sky positions of the multiple images of the source that is removed in (i) with the model-predicted sky positions of the multiple images of the same source derived in (iii).
    \item We repeat this procedure $R$ times using multiple images of $R$ different sources.
\end{enumerate}

\begin{figure}[t]
    \centering
    \includegraphics[width=8cm,clip]{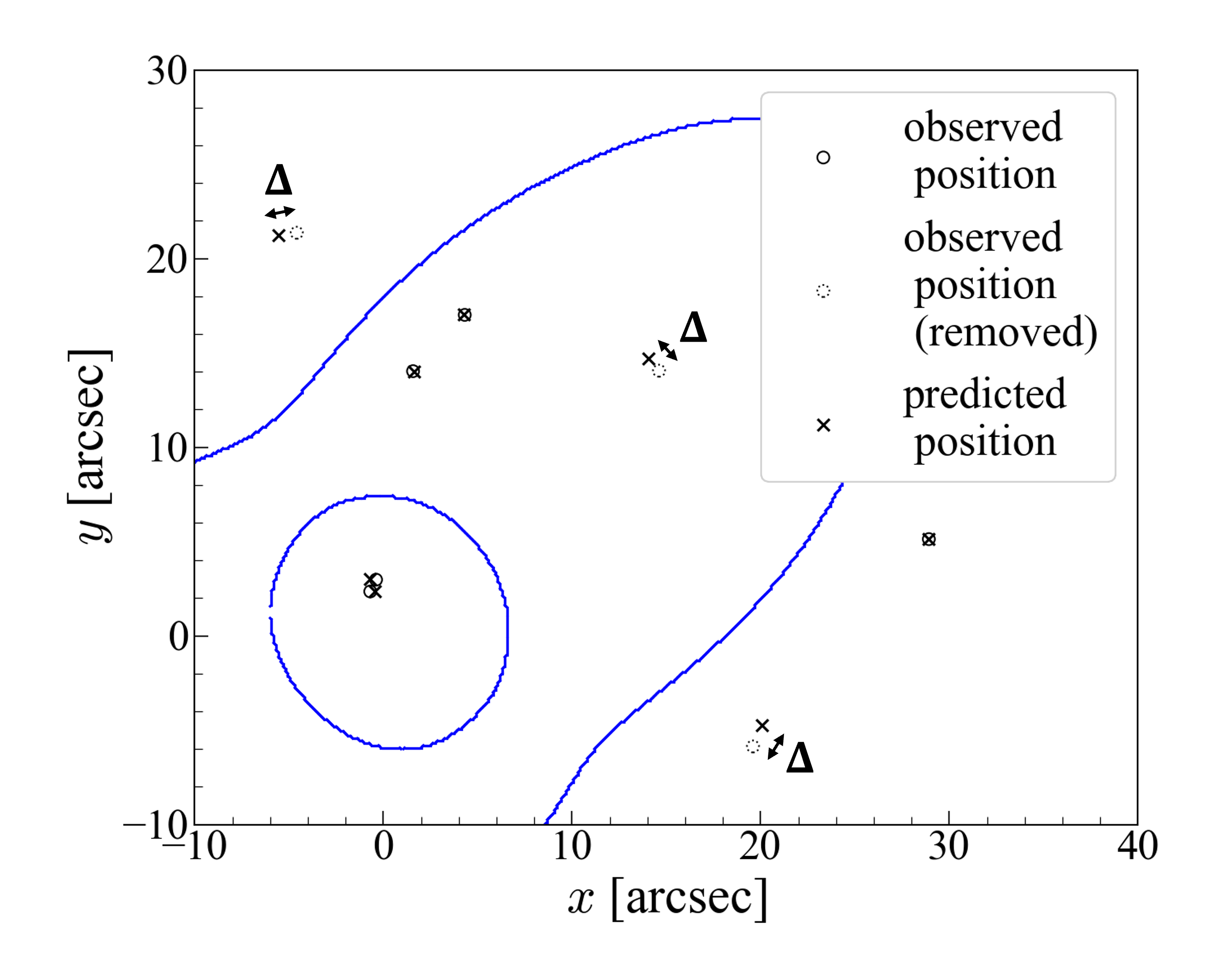}
    \caption{Explanation of the Jackknife method. Solid circles represent observed multiple images and the three dotted circles denote multiple images of a removed source. Crosses show positions of multiple images predicted by the mass model that is derived using the filled circles as constraints. Positional differences between dotted circles and crosses $\Delta=\sqrt{\Delta x^2+\Delta y^2}$ are used for the Jackknife method. Solid lines show critical curves.}
    \label{fig:Jack_exp}
\end{figure}

We denote the positional differences in the image plane for each multiple image derived in (iv) as ($\Delta x$, $\Delta y$), and  its length as $\Delta=\sqrt{\Delta x^2+\Delta y^2}$. We analyze the ratio $\Delta x/\sigma, \Delta y/\sigma$ to evaluate the model performance, where $\sigma$ is the assumed positional error.

If the model is correct, $\Delta$ typically matches $\sigma$ because of the high predictive power. In this case, the distribution of $\Delta x/\sigma, \Delta y/\sigma$ is expected to follow the Gaussian distribution with the standard deviation of 1.
On the other hand, if the model is incorrect and overfitted, $\Delta$ is expected to be greater than $\sigma$ because of the low predictive power. In this case, the distribution of $\Delta x/\sigma, \Delta y/\sigma$ is expected to deviate from the Gaussian distribution with the standard deviation of 1. 

Fig.~\ref{fig:Jack_exp} shows an example of the Jackknife method. 
In this case, $\Delta$ is larger than $\sigma$, indicating an incorrect model.

\section{Test with simulations}
\label{sec:simulations}
\subsection{Description of simulations}
We test the Jackknife method using simulated lensing data generated with {\sc glafic} \citep{2010PASJ...62.1017O,2021PASP..133g4504O}
to verify its effectiveness of the Jackknife method in the strong lensing analysis.

\begingroup 
    \setlength{\tabcolsep}{2pt} 
    \renewcommand{\arraystretch}{1.5} 
    \begin{table}[t]
    \centering
    \begin{tabular}{cccccc}
        Model & $z_{\rm{l}}$ & M$\ (10^{14}\ h^{-1} \ M_\odot)$ & $e$   & $\theta_e\ (\rm{deg})$ & $c$ \\
        \hline 
        \texttt{anfw} & 0.6 & $6.0\times10^{14}$ & 0.5 & $-45$   & 8 \\
        \hline
        \\
        & $z_{\rm{l}}$ & $z_{s, \rm{fid}}$ & $\gamma$   & $\theta_\gamma$ \\
        \hline
        \texttt{pert} & 0.6 & 2.0 & 0.05 & 60  \\
        \hline 
        \\
    \end{tabular}
    \caption{Lens model components for the input model}
    \label{Table:Lens_input}
    \end{table}
\endgroup

We first construct an input lens model for the simulation. 
Table \ref{Table:Lens_input} shows the lens model.
The input lens model includes one dark matter halo with the Navarro-Frenk-White \cite[NFW;][]{1997ApJ...490..493N} density profile, corresponding to {\texttt{anfw}} in {\sc glafic}. The NFW density profile includes a mass $M$, a sky position $x$ and $y$, an ellipticity $e$, a position angle $\theta_e$, and a concentration parameter $c$. We also add an external shear ({\texttt{pert}} in {\sc glafic}) that is a second order ($m = 2, n = 2$) of the external perturbation defined as

\begin{equation}
    \phi=-\frac{\gamma}{m}r^n\cos{m \left( \theta-\theta_{\gamma}-\frac{\pi}{2}\right)}.
    \label{external_perturbation}
\end{equation}
The external shear includes a position angle $\theta_\gamma$ and a constant tidal shear $\gamma$. 
The amplitudes of the perturbations are defined for a fiducial source redshift $z_{\rm{s}, \rm{fid}}$ and are scaled with the source redshift assuming the perturbation comes from the structure at the cluster redshift $z_{\rm{l}}$.

Next, we calculate the positions $(x, y)$ of multiple images using the input model. 
To generate mock data, we add 100 realizations of the Gaussian $\mathcal{N}(0, \sigma_{\rm{eff}})$ to the positions, resulting in 100 mock datasets. 
Here, $\mathcal{N}(0, \sigma_{\rm{eff}})$ is the Gaussian distribution having a mean value of 0 and the standard deviation of $\sigma_{\rm{eff}}$.
The value of $\sigma_{\rm eff}$ was chosen to match the typical RMS $\Delta_{\rm{RMS}}$ defined as

\begin{equation}
    \Delta_{\rm{RMS}} = \sqrt{\frac{1}{P}\sum_{i=1}^{P}|\bm{\theta}_i^{\rm{obs}}-\bm{\theta}_i^{\rm{model}}(a_1,\cdots, a_Q)|^2}.
    \label{chi^2}
\end{equation}
We can assume $\chi^2/\nu = P\Delta_{\rm{RMS}}^2/\nu{\sigma_{\rm{eff}}}^2\approx 1$ if the fitting model is valid. 
Thus we have $\sigma_{\rm{eff}}\approx\sqrt{P/\nu}\Delta_{\rm{RMS}}\approx\Delta_{\rm{RMS}}$.
As is summarized e.g., in \cite{2023ApJ...951..140C}, the typical RMS value in the strong lensing analysis of massive clusters is roughly $0\farcs4-0\farcs5$. 
Thus we assume $\sigma_{\rm eff}=0\farcs4$ throughout the simulations.

We optimize both correct and incorrect models using the mock multiple image positions derived above, and obtain the $\chi^2$-distribution with 100 $\chi^2$ values from the 100 mock data.
Here the correct model has exactly the same lens components as the input model.
In contrast, the incorrect model assumes 3 additional multipole perturbations ($m=3$, $4$, $5$, $n=2$) defined in Eq.~(\ref{external_perturbation}) in addition to the lens components used in the input model.
In this simulation set up, we note that, when the multipole perturbations are small, the incorrect model effectively reduces to the correct model.
The correct model is optimized with the original value of $\sigma_{\rm{eff}}$, i.e., $\sigma=\sigma_{\rm{eff}}$. 
The incorrect model is optimized assuming the positional error of $\sigma=\sigma'\neq \sigma_{\rm{eff}}$, where $\sigma'$ is determined such that reduced chi-square $\chi^2/\nu \approx 1$.

More specifically, for obtaining $\sigma'$, we first optimize the incorrect model assuming $\sigma_{\rm{eff}}$ as the positional error. In this case, the $\chi^2$-distribution is expected to be deviated from the $\chi^2$-distribution with a true DoF $\nu$. The mean value of the $\chi^2$-distribution, $\langle \chi^2(\sigma_{\rm{eff}}) \rangle$, is represented as
\begin{equation}
    \langle \chi^2(\sigma_{\rm{eff}})\rangle = \frac{1}{{\sigma_{\rm{eff}}}^2} \Big\langle \sum_{i=1}^{P}|\bm{\theta}_i^{\rm{obs}}-\bm{\theta}_i^{\rm{model}}|^2 \Big\rangle.
    \label{chi_dis_mean}
\end{equation} 
When we assume $\sigma^{\prime}$ as the positional error to obtain the $\chi^2$-distribution with the true DoF $\nu$, the mean value of $\chi^2$-distribution corresponds to the true DoF $\nu$. Thus
\begin{equation}
    \nu = \frac{1}{{\sigma^{\prime}}^2} \Big\langle \sum_{i=1}^{P}|\bm{\theta}_i^{\rm{obs}}-\bm{\theta}_i^{\rm{model}}|^2 \Big\rangle. 
    \label{chi_dis_mean_2}
\end{equation} 
With Eqs.~(\ref{chi_dis_mean}) and (\ref{chi_dis_mean_2}), we can rescale $\sigma_{\rm{eff}}$ to obtain $\sigma'$, ensuring that it follows the $\chi^2$-distribution for the true DoF.
Specifically, $\sigma'$ is given as
\begin{equation}
    \sigma'= \sqrt{{\frac{\langle \chi^2(\sigma_{\rm{eff}}) \rangle}{\nu}}} \sigma_{\rm{eff}}.
\end{equation}
Therefore if we optimize the incorrect model with $\sigma'$, the value of the reduced chi-square $\chi^2/\nu$ is $1$ on average. 
This means that we cannot distinguish between the correct and incorrect models based on the reduced chi-square, since both models have $\chi^2/\nu \approx 1$. The purpose of our mock data analysis to check whether the Jackknife method can distinguish the correct model from the incorrect model.

We apply the Jackknife method to these two models 
to obtain distributions of $\Delta x/\sigma_{\rm{eff}}, \Delta y/\sigma_{\rm{eff}}$ and $\Delta x/\sigma^{\prime}, \Delta y/\sigma^{\prime}$.
We call this distribution the Jackknife distribution.
In our analysis, we use several set-ups with 5, 6, 7, 10, and 20 sources to check the dependence of our results on the number of sources.
The source redshifts $z_{\rm s}$ are randomly set in the range $z_{\mathrm{s}}=2.0-5.0$ and are also included as model parameters with informative Gaussian priors with an error of 0.1, 
corresponding to the case where we have photometric redshifts, so that the model can overfit the data.

\begingroup 
    \setlength{\tabcolsep}{6pt} 
    \renewcommand{\arraystretch}{1.5} 
    \begin{table*}[t]
    \centering
    \begin{tabular}{cccc|ccccc} 
        \multicolumn{4}{c}{input model} & \multicolumn{4}{c}{fitting model}\\ 
        \hline
        Source  & Image &
        \begin{tabular}{c}
             Realization\\
             of $\mathcal{N}(0, \sigma_{\rm{eff}})$
        \end{tabular}
        & 
        \begin{tabular}{c}
             Realization\\
             of source positions
        \end{tabular} 
        & Model & Constraint & Parameter & DoF \\
        \hline 
        5 & 15 & 100 & 2 & Correct & 35 & 23 & 12\\
        &&&& Incorrect & 35 & 29 & 6\\
        &&&& 
        \begin{tabular}{c}
             Incorrect\\
             (2 haloes)
        \end{tabular} & 35 & 29 & 6\\
        \hline
        6 & 18 & 100 & 4 & Correct & 42 & 26 & 16\\
        &&&& Incorrect & 42 & 32 & 10\\
        \hline
        7 & 21 & 100 & 4 & Correct & 49 & 29 & 20\\
        &&&& Incorrect & 49 & 35 & 14\\
        \hline
        10 & 30 & 100 & 2 & Correct & 70 & 38 & 32\\
        &&&& Incorrect & 70 & 44 & 26\\
        \hline
        20 & 60 & 100 & 1 & Correct & 140 & 68 & 72\\
        &&&& Incorrect & 140 & 74 & 66\\
        \hline   
        \\
    \end{tabular}
    
    \caption{Setting for input and fitting models}
    \label{Table:Setting_input_fitting}
    \end{table*}
\endgroup

The details of the input and fitting model (both correct and incorrect models) settings are summarized in Table \ref{Table:Setting_input_fitting}.
For each number of sources, we consider multiple realizations of source positions, and derive the Jackknife distribution for each realization. 'Constraint' refers to the total number of observational constraints, namely the positions and redshifts of the multiple images. 
'Parameter' refers to the number of parameters used for the model fitting, such as source redshifts, source positions, and lens model parameters.
'DoF' is calculated as 'Constraint' minus 'Parameter'.
For the incorrect model with 5 sources, we also conduct the mock analysis assuming 2 dark matter haloes, as indicated by 5 (2 haloes).

\subsection{Result}
\begin{figure*}[h]
  \centering
  \includegraphics[width=14cm,clip]{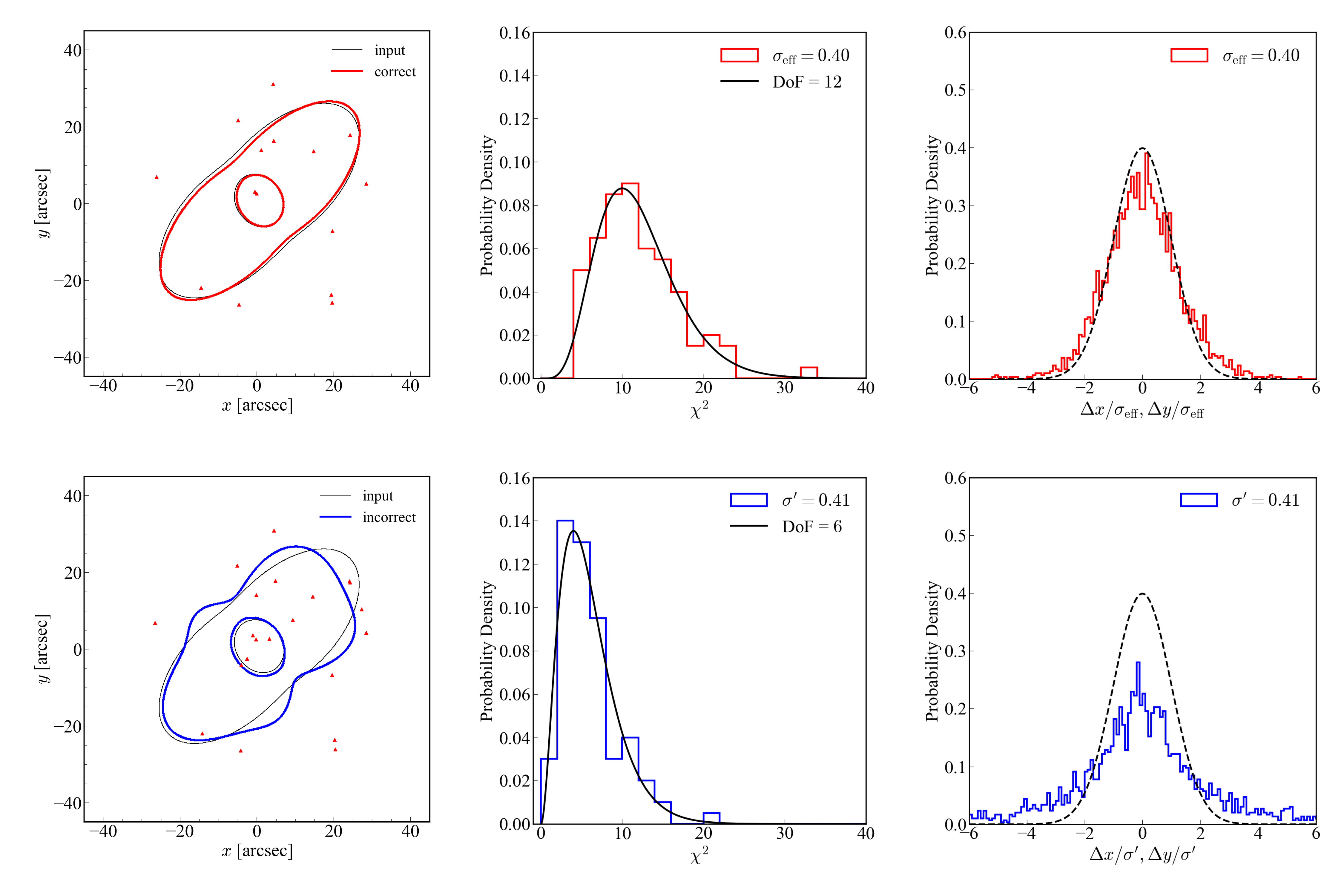}
  \caption{
  Result of the test of the Jackknife method with simulations for 5 sources. 
  Upper panels show the result of the correct model and lower panels show the result of the incorrect model.
  Left panels show critical curves of the input, correct, and incorrect models. Mock multiple images used for the analysis are indicated by filled triangles.
  Middle panels show the $\chi^2$-distribution for the correct and incorrect models. The correct model is optimized assuming the input positional error, $\sigma_{\rm{eff}}$. 
  The incorrect model is optimized assuming the modified positional error $\sigma'$ so as to reproduce the reduced chi-square of $1$ on average.
  Right panels show the Jackknife distribution (see the text for the definition) for the correct and incorrect models. 
  The dotted line represents the Gaussian distribution $\mathcal{N}(0, 1)$.
 Since we use source-plane fitting to reduce computation time, the number of predicted images can differ  from the number of observed images. For the incorrect model, some extra images are predicted by the model. Some of the additional images appear near the halo centers are usually faint and are difficult to be identified in observations.}
  \label{fig:Result_5_points}
\end{figure*}

\begin{figure*}[h]
  \centering
  \includegraphics[width=14cm,clip]{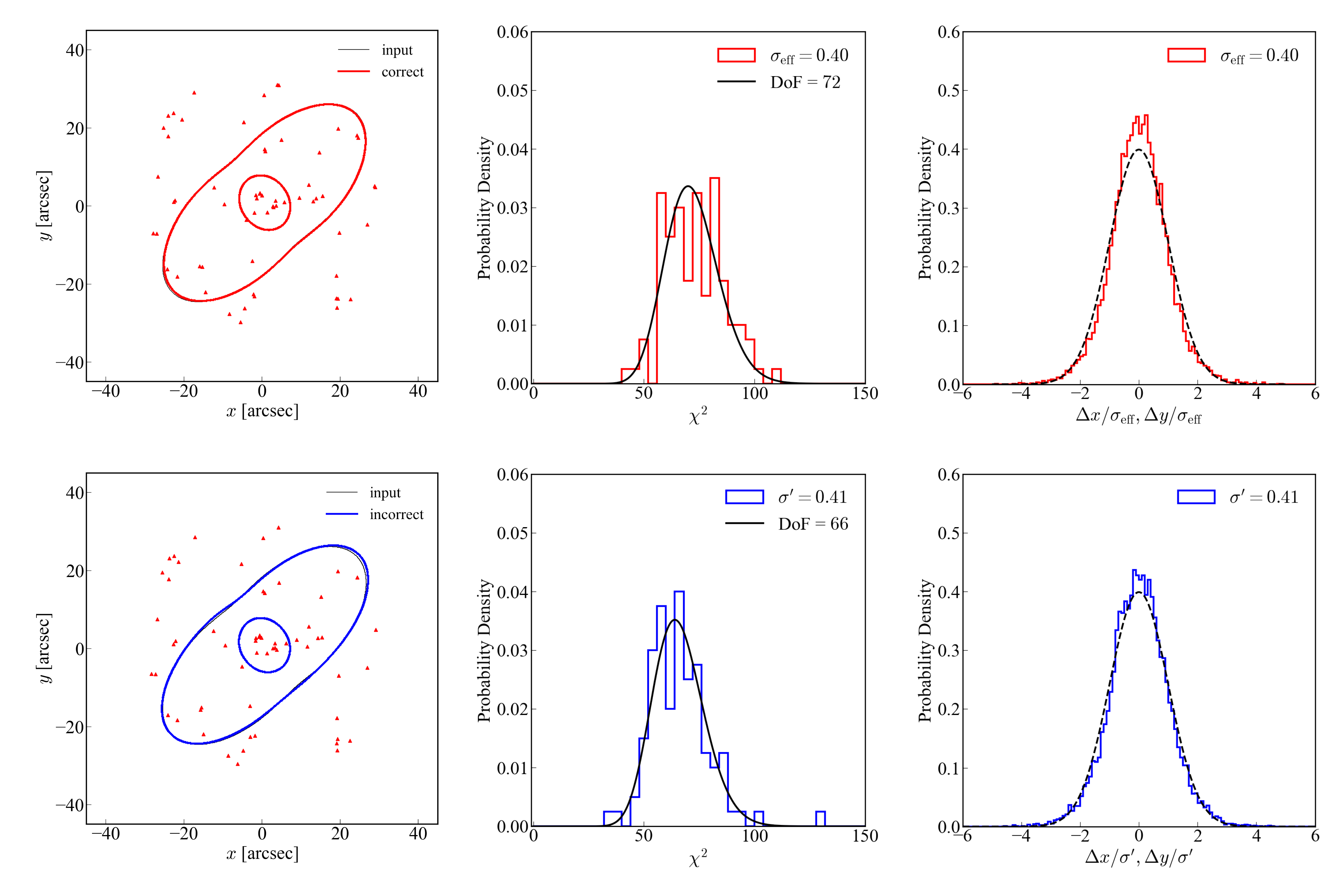}
  \caption{
  Same as Fig.~\ref{fig:Result_5_points}, but for 20 sources.
  }
  \label{fig:Result_20_points}
\end{figure*}
Fig.~\ref{fig:Result_5_points} shows the result of the simulation using 5 sources. The correct model follows the $\chi^2$ -distribution with the true DoF.
The incorrect model is optimized with $\sigma'$ such that $\chi^2/\nu$ is $1$ on average. Thus, the incorrect model also follows $\chi^2$ -distribution with the true DoF. 
Both these models cannot be distinguished from the reduced chi-square, while the example critical curves shown in Fig.~\ref{fig:Result_5_points} suggests that the incorrect model significantly deviates from the input model.
 
We find that the Jackknife method indeed reveals the difference between the correct and incorrect models.
While $\Delta x/\sigma_{\rm{eff}}, \Delta y/\sigma_{\rm{eff}}$ for the correct model  follows the Gaussian distribution $\mathcal{N}(0, 1)$, $\Delta x/\sigma^{\prime},\Delta y/\sigma^{\prime}$ for the incorrect model deviates from Gaussian $\mathcal{N}(0, 1)$.
Specifically, the standard deviation of $\Delta x/\sigma_{\rm{eff}}, \Delta y/\sigma_{\rm{eff}}$ for the correct model of 1.24 is close to 1, while the standard deviation of $\Delta x/\sigma^{\prime},\Delta y/\sigma^{\prime}$ for the incorrect model of 2.28 is much larger than 1.
This indicates that the Jackknife method successfully identifies the incorrect model as overfitting.
We confirm that the Jackknife distribution for the incorrect model of 5 sources remains consistent for different realizations of source positions (see Appendix~\ref{ap:jackknife_mcmc}).

Fig.~\ref{fig:Result_20_points} shows result of the simulation using 20 sources.
In this case, the standard deviation of the Jackknife distribution is 1.02 for the correct model, and 1.14 for the incorrect model, which means that both models follow the Gaussian $\mathcal{N}(0, 1)$ and there is almost no difference in the Jackknife distribution between the two models.
As seen in the critical curves of the correct and incorrect models in Fig.~\ref{fig:Result_20_points}, the best-fitting model for the incorrect model is almost identical to that for the correct model. 
This is because it is more difficult for the incorrect model to overfit due to the increased number of multiple images.
The increased number of constraints limits the flexibility of the incorrect model and prevents overfitting.
The Jackknife distribution with no difference between the two models is consistent with this result.

\section{Demonstration with actual observations}
\label{sec:demonstration}
\begin{figure*}[t]
  \centering
  \includegraphics[width=16cm,clip]{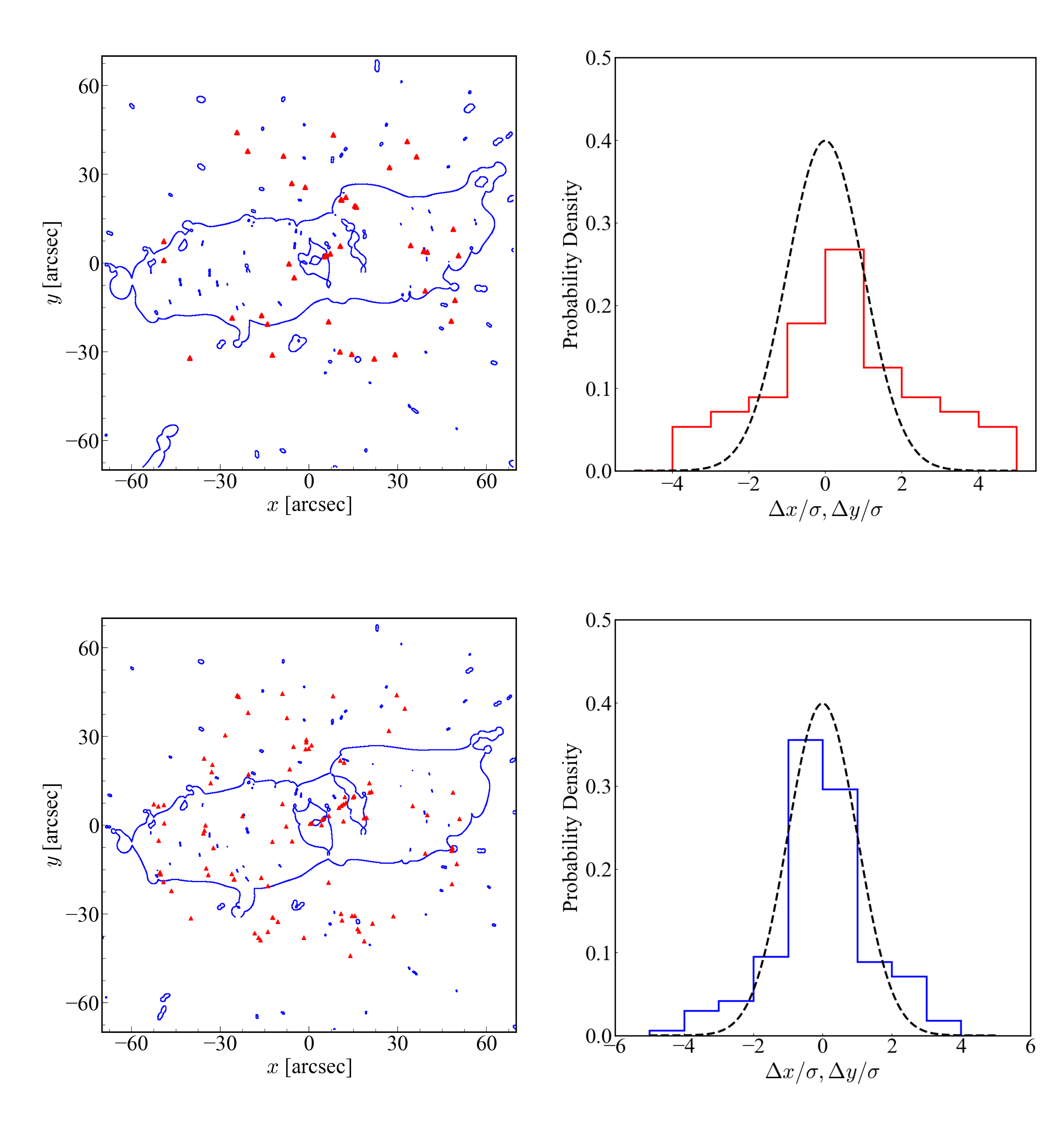}
  \caption{Result of applying the Jackknife method to the actual observations of MACS0647. 
  Upper panels show the result for the mass model based on the HST observation and lower panels show the results of the JWST observation.
  Left panels show critical curves ({\it solid lines}) and multiple images ({\it filled triangles}) for HST and JWST observations.
  Right panels show the results of the Jackknife distribution for the HST and JWST observations. The dotted line represents Gaussian $\mathcal{N}(0, 1)$.
  }
  \label{fig:Result_MACS0647}
\end{figure*}

As a demonstration using actual observations, we apply the Jackknife method 
to evaluate the predictive power of the mass models
optimized by multiple images observed by the Hubble Space Telescope (HST) and the James Webb Space Telescope (JWST) for the distant galaxy cluster MACS0647 at $z=0.584$.
This lensing cluster is well known for hosting MACS0647-JD in its lensed background galaxy, a distant galaxy at $z_{\rm s}=10.17$, making it one of the earliest known galaxies \citep{2013ApJ...762...32C, 2023ApJ...949L..34H}. The detailed information of the mass models of the galaxy cluster is given in Appendix~\ref{ap:mass_model}.

The results are shown in Fig.~\ref{fig:Result_MACS0647}.
The Jackknife distribution by the HST deviates from the Gaussian distribution $\mathcal{N}(0, 1)$ with the standard deviation of 4.03, indicating that the model is incorrect and overfitted.
In contrast, the Jackknife distribution by the JWST is closer to the Gaussian distribution $\mathcal{N}(0, 1)$ with the standard deviation of 1.72, suggesting that it is relatively a well-constrained and accurate mass model.
This difference may originate from the higher sensitivity of the JWST due to the larger aperture, which enables the detection of a larger number of multiple images.
In fact, the number of multiple images observed with the HST is 31, while it increases to 86 with the JWST (see Appendix~\ref{ap:mass_model}).
The larger sample size of multiple images prevents the mass model from overfitting by constraining the model parameters more tightly.

\section{Discussion}
\label{sec:discussion}

\subsection{Validating the estimated error of the physical quantities by MCMC}

\begin{figure}[t]
  \centering
  \includegraphics[width=8cm,clip]{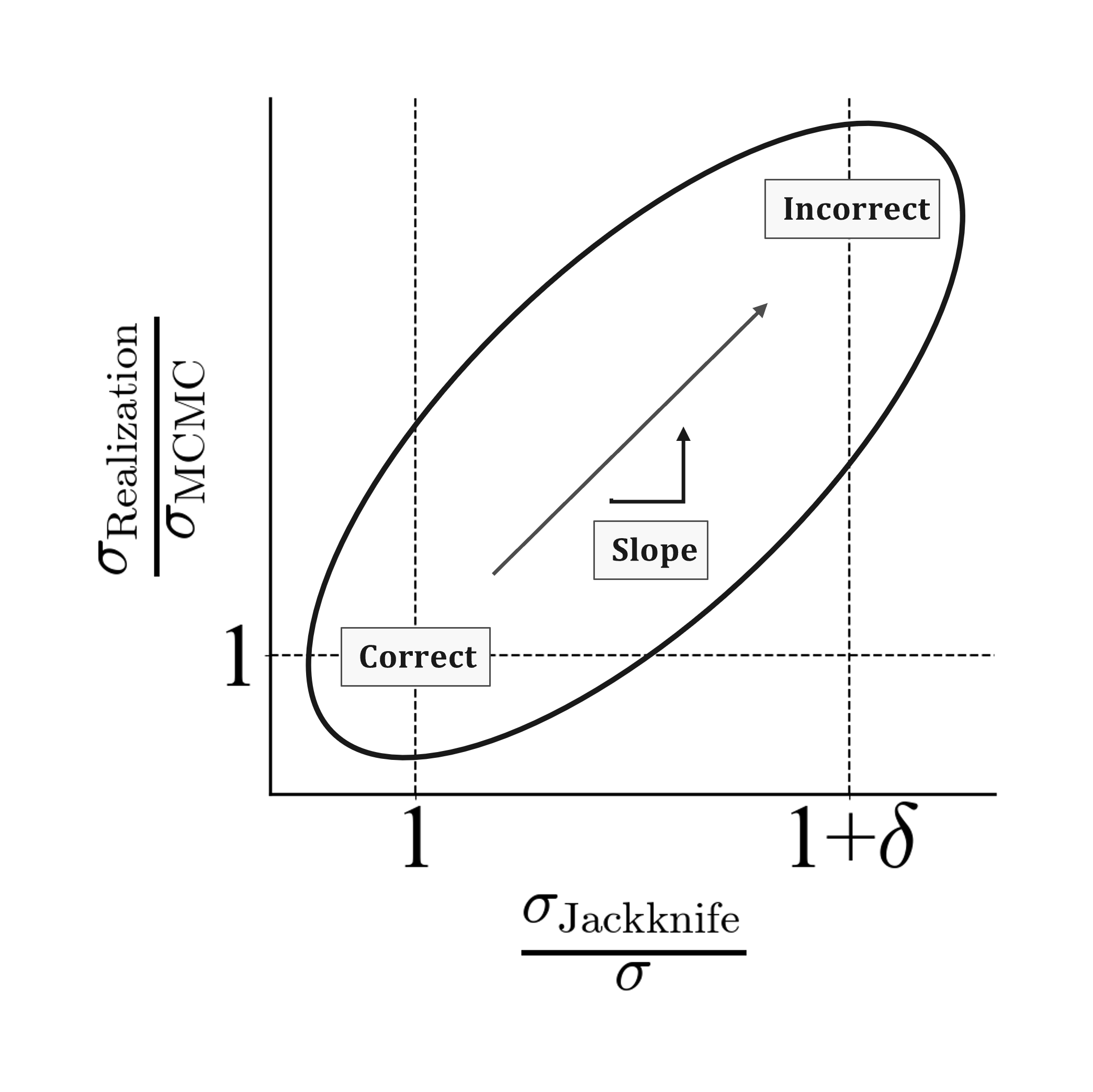}
  \caption{Schematic diagram showing an expected trend between $\sigma_{\rm{Jackknife}}/\sigma$ and $\sigma_{\rm{Realization}}/\sigma_{\rm{MCMC}}$. 
  We define $\delta$ as the difference of $\sigma_{\rm{Jackknife}}/\sigma$
  between correct and incorrect models. 
  The slope is defined by the gradient when plotting $\sigma_{\rm{Jackknife}}/\sigma$ in the horizontal axis and $\sigma_{\rm{Realization}}/\sigma_{\rm{MCMC}}$ in the vertical axis. The ellipse shows the expected overall trend between $\sigma_{\rm{Jackknife}}/\sigma$ and $\sigma_{\rm{Realization}}/\sigma_{\rm{MCMC}}$, including the scatter.}
  \label{fig:Trend}
\end{figure}

\begin{figure*}[t]
  \centering
  \includegraphics[width=16cm,clip]{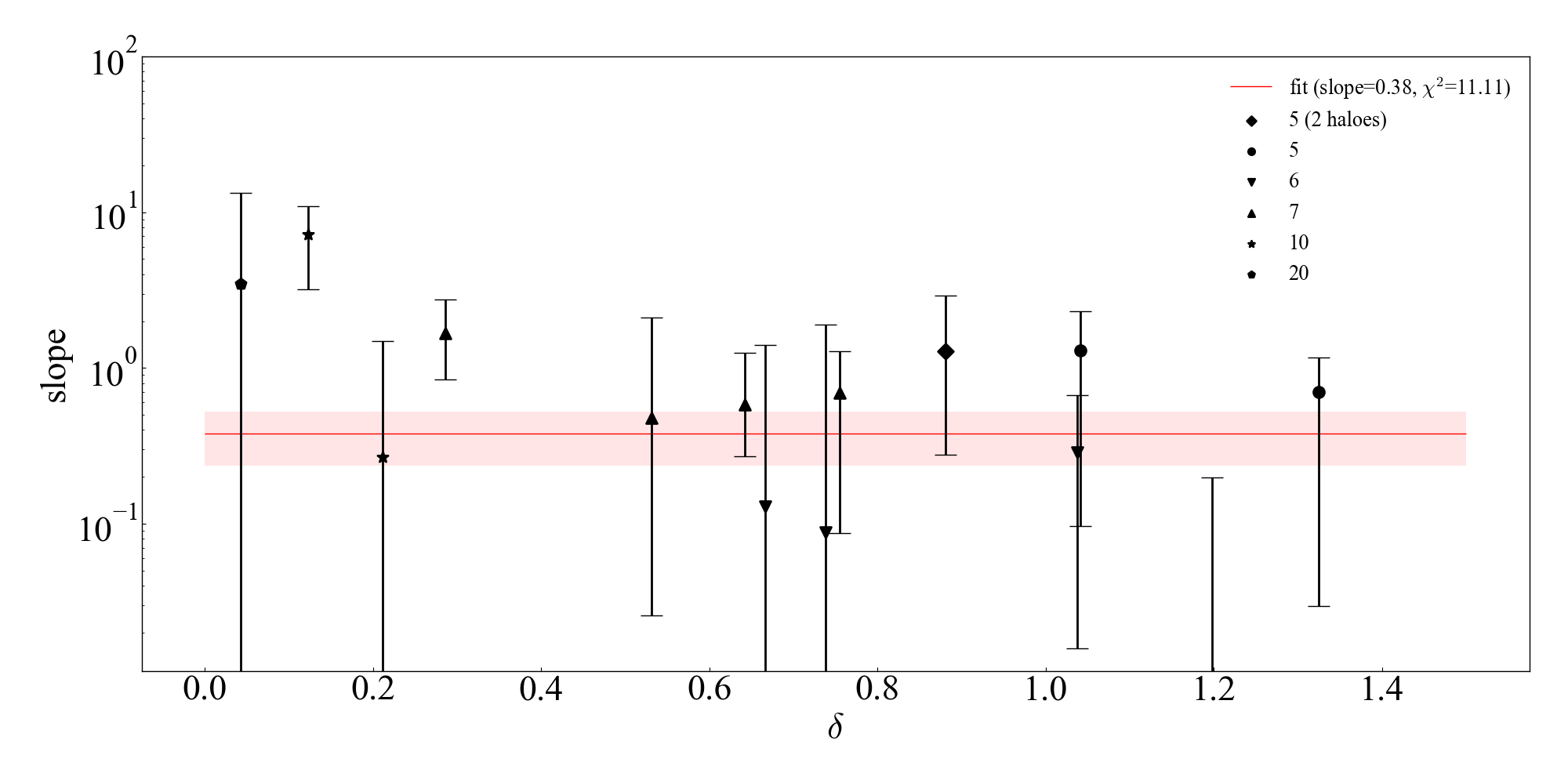}
  \caption{The slope defined in Fig.~\ref{fig:Trend} for magnifications as a function of the degree of overfitting $\delta$ that is also defined in Fig.~\ref{fig:Trend}.
  In order to derive the slope for a wide range in $\delta$, we conduct the mock strong lens data analysis with set-ups summarized in Table~\ref{Table:Setting_input_fitting}.
  Filled triangles denote median values for individual set-ups, and the errors represent 16 and 84 percentiles.
  The red line is a best-fitting median of the slope, assuming a constant slope. 
  The shaded region shows a 68${\%}$ confidence interval.}
  \label{fig:Mag_delta_slope}
\end{figure*}

\begin{figure*}[t]
  \centering
  \includegraphics[width=16cm,clip]{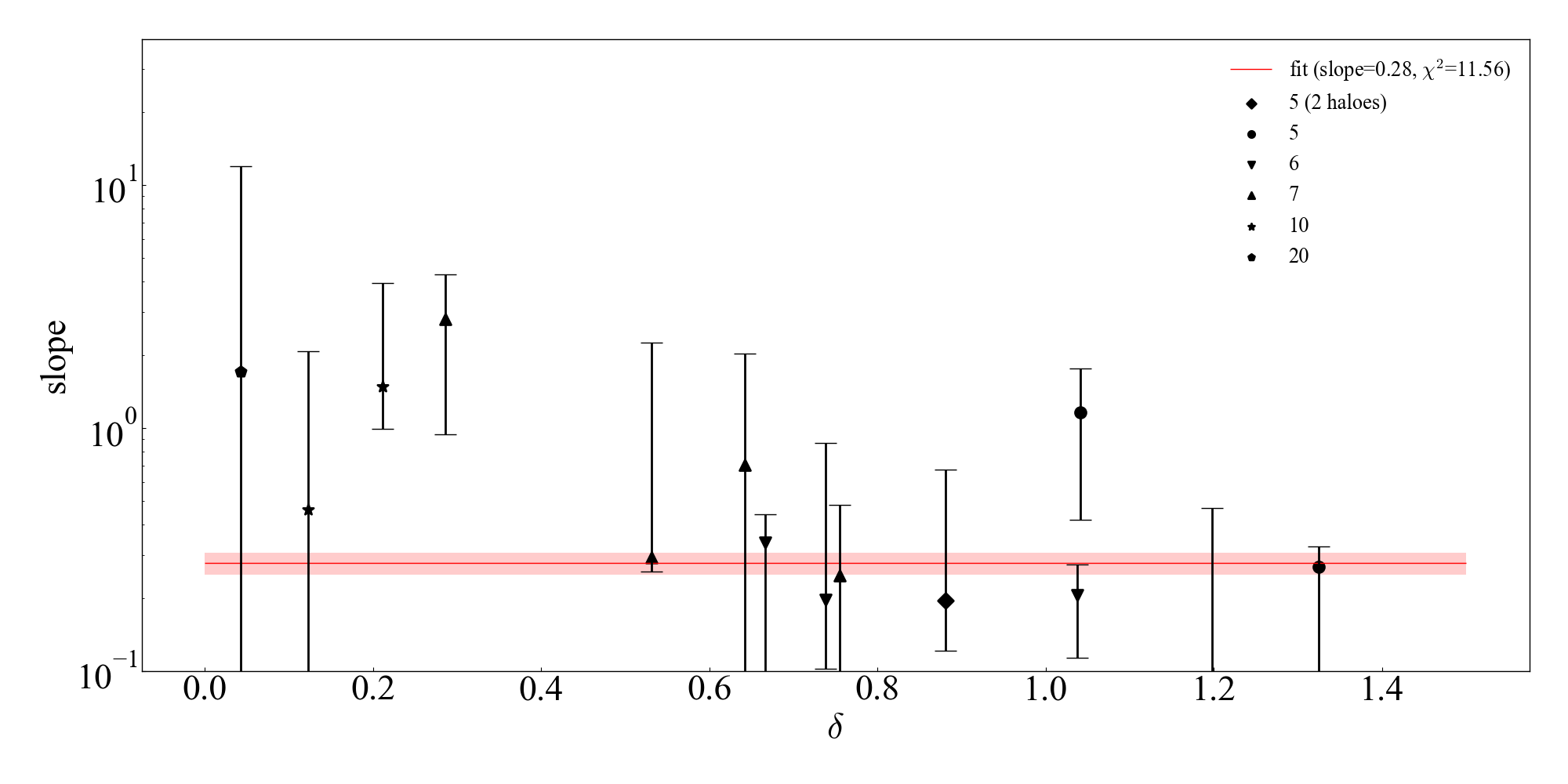}
  \caption{
  Same as Fig.~\ref{fig:Mag_delta_slope}, but for time delays.
  }
  \label{fig:TimeDelay_delta_slope}
\end{figure*}

The correct model in our definition is expected to predict physical quantities accurately.
In the strong lensing analysis, the physical quantities that we are interested in are magnifications and time delays, which provide important information about distant galaxies and supernovae as well as cosmology. 

In the strong lensing analysis, the MCMC is used for estimating errors of the physical quantities such as magnifications and time delays.
In this paper, we introduce a new approach to evaluate the validity of errors estimated by the MCMC using the Jackknife method.

As a preparation, we define $\sigma_{\rm{Jackknife}}$, $\sigma_{\rm{Realization}}$ and $\sigma_{\rm{MCMC}}$.
\begin{itemize}
    \item $\sigma_{\rm{Jackknife}}$: The width of the distribution $\Delta x, \Delta y$. 
    By comparing it with the assumed positional error $\sigma$, the ratio $\sigma_{\rm{Jackknife}}/\sigma$ serves as an indicator of model overfitting.
    \item $\sigma_{\rm{realization}}$: The width of the distribution of a physical quantity across all realizations, representing the true error of the physical quantity.
    \item $\sigma_{\rm{MCMC}}$: The width of the MCMC sample distribution of a physical quantity for a single realization, representing the estimated error of the physical quantity. The ratio $\sigma_{\rm{realization}}/\sigma_{\rm{MCMC}}$ is the ratio of the estimation error of the physical quantity to the true error, and serves as an indicator of the degree of the underestimation of the error on the physical quantity by the MCMC.
\end{itemize}

Since the correct model has the high predictive power, $\sigma_{\rm{Jackknife}}/\sigma$ is close to 1 as we see in the analysis with simulations.
In addition, the estimated error of a physical quantity is expected to be close to the true error if the model is correct. 
Thus $\sigma_{\rm{realization}}/\sigma_{\rm{MCMC}}$ is also close to 1. On the contrary, the predictive power of the incorrect model is low, and $\sigma_{\rm{Jackknife}}/\sigma$ is larger than $1$ as we see in the analysis with simulations.
In particular, when the model is overfitted, we expect that the error of a physical quantity estimated by the MCMC is smaller than the true error due to biased estimates of a physical quantity, which means that $\sigma_{\rm{realization}}/\sigma_{\rm{MCMC}}$ is larger than 1.
Hence we expect that the relation between $\sigma_{\rm{Jackknife}}/\sigma$ and $\sigma_{\rm{realization}}/\sigma_{\rm{MCMC}}$ follows the trend shown in Fig.~\ref{fig:Trend}. 

Once this trend is established, we can validate the statistical error of a physical quantity obtained by the MCMC, $\sigma_{\rm{MCMC}}$, taking advantage of the relation between $\sigma_{\rm{Jackknife}}/\sigma$ and $\sigma_{\rm{realization}}/\sigma_{\rm{MCMC}}$. 
In this case, once we obtain $\sigma_{\rm{Jackknife}}/\sigma$ from observations, we can infer $\sigma_{\rm{realization}}/\sigma_{\rm{MCMC}}$. Since we know the value of $\sigma_{\rm{MCMC}}$, we can estimate the true error $\sigma_{\rm{realization}}$ from the inferred ratio. 
Thus, even if $\sigma_{\rm{MCMC}}$ is underestimated, we can obtain the true error of a physical quantity using the Jackknife method.

We perform the mock strong lens data analysis of the Jackknife method with various set-ups summarized in Table~\ref{Table:Setting_input_fitting} to check whether this trend exists. 
We also change the realization of source positions for 5, 6, 7, and 10 sources and compute the slope. Results of the Jackknife distributions for each set-ups are shown in Appendix~\ref{ap:jackknife_mcmc}.
Note that the calculations of the Jackknife distributions for 5 and 20 sources to constrain the slope are performed independently from those used to compute the Jackknife distributions in Fig.~\ref{fig:Result_5_points} and Fig.~\ref{fig:Result_20_points}. While they have same realizations of source positions, the realizations of  $\mathcal{N}(0, \sigma_{\rm{eff}})$ are different, resulting in slightly different Jackknife distributions.
As shown in Fig.~\ref{fig:Mag_delta_slope}, 
we found a subtle trend of a positive slope in both magnifications and time delays.
However, errors on the slope are large, especially with 20 sources. 
The reason is that the difference of $\sigma_{\rm{Jackknife}}/\sigma$ between correct and incorrect models ($\delta$ in Fig.~\ref{fig:Trend}) is very small with 20 sources. This causes the slope to become very steep and the distribution to scatter. 
By combining results for all the set-ups, we find that the slope is on average positive at 2$\sigma$ for both magnifications and time delays. More analysis is needed to fully establish the trend shown in Fig.~\ref{fig:Trend}. 
More simulations with different set-ups may help understand the origin of the scatter of the slope, which we leave for future work.

\subsection{Simulation with cluster member galaxies}

\begin{figure*}[t]
  \centering
  \includegraphics[width=14cm,clip]{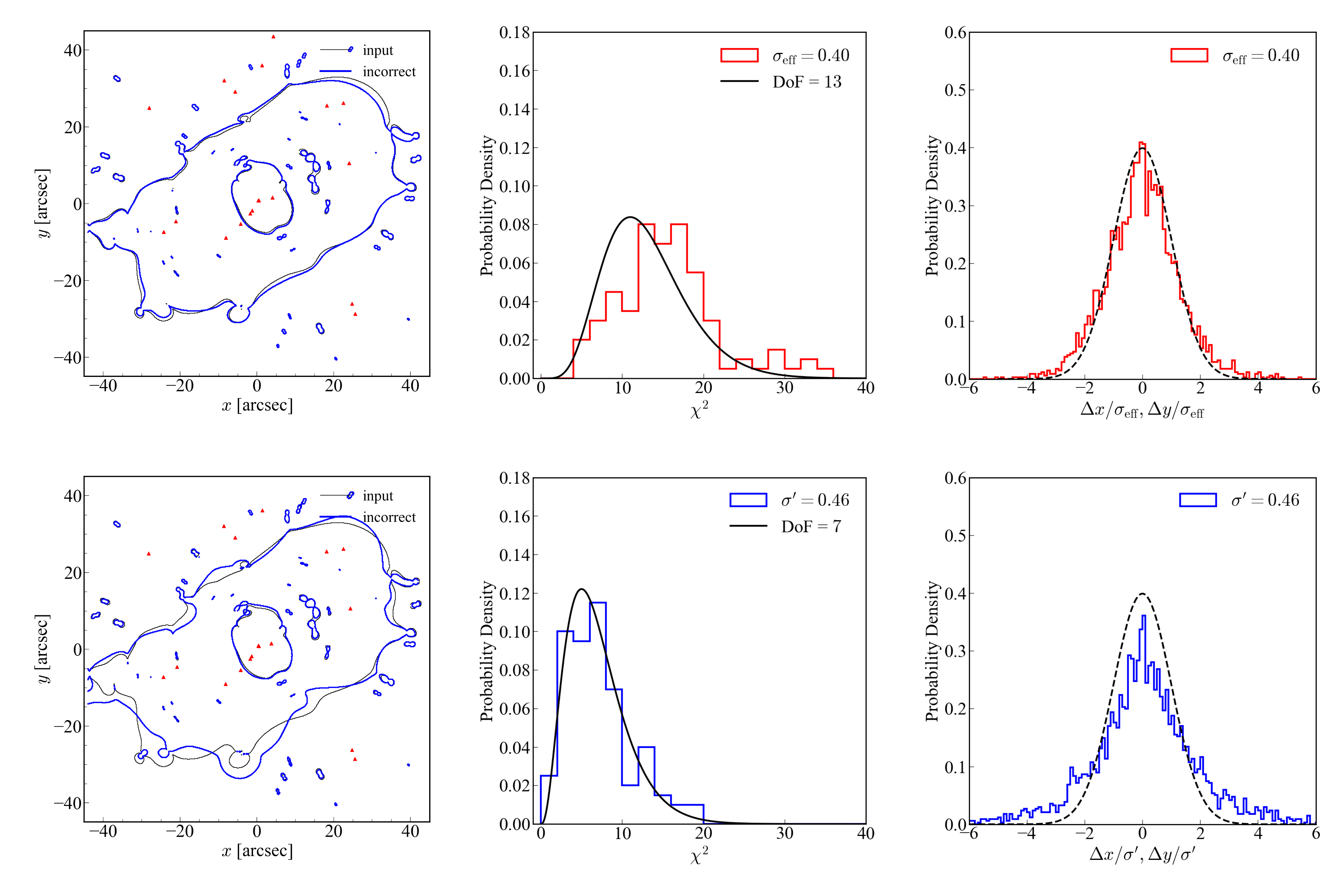}
  \caption{Result of the test of the Jackknife method for a more realistic model with 5 sources using the member galaxies of MACS0647.}
  \label{fig:real_model}
\end{figure*}

It is important to test the validity of the Jackknife method using more realistic lens models. In our previous simulations, we consider the simple lens model without any member galaxies, as seen from the smooth shapes of the critical curves. As an additional test, we use a more complex model that adds the member galaxies of MACS0647 to the NFW plus external shear model considered in Sect.~\ref{sec:simulations}, and test whether the Jackknife method can still distinguish an incorrect model. We use the same set-up as in Table~\ref{Table:Lens_input} with 5 point sources, including the member galaxies of MACS0647. 
We include three parameters modeling the scaling relations of cluster member galaxies as commonly done in cluster strong lens mass modeling \citep[e.g.,][]{2016ApJ...819..114K}.

The result is shown in Fig.~\ref{fig:real_model}. We find that the critical curve of the incorrect model significantly deviates from that of the input model. By applying the Jackknife method, we find a significant deviation of the Jackknife distribution from $\mathcal{N}(0, 1)$ with the standard deviation of 1.65, indicating that the method successfully identifies the incorrect model as overfitting. This suggests that the Jackknife method remains effective even for realistic and complex lens models.

\section{Conclusion}
\label{sec:conclusion}
Cluster strong lensing serves as a powerful tool to constrain the mass distribution of a galaxy cluster. 
The higher sensitivity of telescopes allows us to observe an increased number of multiple images to strengthen the constraint on the mass model. 
However, it is very important to validate the best-fitting mass model 
since the accuracy of the mass model directly translates into the accuracy of physical quantities derived from the mass model such as magnifications and time delays.
The reduced chi-square, which is commonly used to quantify the validity of fitting, cannot be used here as a validation because of the uncertainty of the positional error arising due to the complexity of the dark matter distribution.
How to validate the strong lens mass model is still an open question.

In this paper, we have proposed the Jackknife method as a new method to validate the mass model by quantifying its predictive power.
In the Jackknife method, we remove the multiple images for one of sources and then optimize the mass model using the rest of multiple images. 
Using the best-fitting mass model, we predict positions of the multiple images of the removed source. 
We then obtain the positional difference between observed and model-predicted multiple images of the removed source. 
Repeating this procedure for different removed sources, we can evaluate the predictive power of the mass model by checking the difference between these positional differences from the Jackknife method and assumed positional errors.

To test the effectiveness of the Jackknife method, we have conducted the analysis on the mock strong lens data to check whether the Jackknife method can distinguish the correct and incorrect, overfitted models, both of which have the correct model and incorrect model with reduced chi-square $\chi^2/\nu \approx 1$.
The Jackknife method has been shown to work well in simulations assuming a simple mass model with a single dark matter halo and 5 sources.
In the case of 20 sources, 
no significant differences between input and fitted models in the Jackknife distribution have been observed.
However, this result is reasonable in the sense that the incorrect model is not overfitted  thanks to the stringent constraints with a large number of multiple images.
We have applied the Jackknife method to strong lens mass models of the galaxy cluster MACS0647 to demonstrate the feasibility.

Additionally, we have explored the potential of the Jackknife method for estimating the accuracy of errors of physical quantities such as magnifications and time delays estimated by the MCMC. 
Considering the predictive power of the mass model,
we expect the trend shown in Fig~\ref{fig:Trend}. We have checked whether this trend exists in our mock strong lens analysis with various set-ups. 
While we find a possible trend in our analysis,
further studies with a wider range of set-ups may help verify the effectiveness more robustly.

While the new Jackknife method proposed in this paper has proven to be effective from the analysis on simulations with relatively simple mass models, it is of great importance to explore the effectiveness and limitation of the Jackknife method with simulations with more realistic set-ups. We leave such explorations for future work.

\section*{Acknowledgments}

We thank Sherry Suyu for discussions. 
We thank an anonymous referee for useful suggestions.
This work was supported by JSPS KAKENHI Grant Numbers JP25H00662, JP25H00672, and JP22K21349. 
This work is based on observations made with the NASA/ESA/CSA James Webb Space Telescope. The data were obtained from the Mikulski Archive for Space Telescopes at the Space Telescope Science Institute, which is operated by the Association of Universities for Research in Astronomy, Inc., under NASA contract NAS 5-03127 for JWST. These observations are associated with program 1433 and 
can be accessed via \href{https://doi.org/10.17909/wpys-ap03}{DOI:    10.17909/wpys-ap03}.

\bibliographystyle{apsrev4-1}

\bibliography{ref}

\begin{appendix}

\section{JWST and HST mass models of MACS0647}
\label{ap:mass_model}

The mass models of the massive galaxy cluster MACS J0647.7+7015 \citep[MACS0647;][]{2007ApJ...661L..33E} at $z=0.584$ are constructed with both the {\it HST} and {\it JWST} data using the {\sc glafic} software \citep{2010PASJ...62.1017O,2021PASP..133g4504O}. The {\it HST} imaging data were obtained as part of the {\it Cluster Lensing and Supernova survey with Hubble} ({\it CLASH}) program \citep{2012ApJS..199...25P}. We largely follow multiple image identifications of \citet{2015ApJ...801...44Z} to construct the {\sc glafic} mass model, which was presented in \citet{2020MNRAS.496.2591O}. The {\it JWST} mass model is newly constructed taking advantage of the {\it JWST} general observer (GO) program 1433 (PI: Dan Coe) data, which are publicly available in the Mikulski Archive for Space Telescopes. 

While some of the new multiple images identified from the {\it JWST} data were presented in \citet{2023ApJ...944L...6M}, we also add a few new multiple image systems based on our analysis. See e.g., \citet{2023ApJ...949L..34H} and \citet{2024ApJ...973....8H} for the detail of the {\it JWST} observations of MACS0647. Spectroscopic identifications of high-redshift galaxies behind MACS0647 are presented in Worku et al. (in prep.).

The multiple images used for the {\it HST} and {\it JWST} mass modeling are summarized in Tables~\ref{Table:MACS0647_1} and \ref{Table:MACS0647_2}. For the {\it HST} mass model, we use 31 multiple images from 11 sources. We include one halo component assuming an NFW profile, member galaxies modeled by pseudo-Jaffe profiles \citep[see e.g.,][for more detail on how to model the effect of member galaxies]{2016ApJ...819..114K}, and include external shear as well as the multipole perturbations with $m=3$ and $4$. Assuming the positional error of $0\farcs4$, the best-fitting {\it HST} mass model has $\chi^2/{\rm DoF}=24.3/20$ \citep{2020MNRAS.496.2591O}. For the {\it JWST} mass model, we use 86 multiple images from 28 sources. We include two halo components assuming an NFW profile, member galaxies modeled by pseudo-Jaffe profiles, and include external shear as well as the multipole perturbations with $m=3$ and $4$. Assuming the positional error of $0\farcs4$, the best-fitting {\it JWST} mass model has $\chi^2/{\rm DoF}=136/95$. 

\begin{deluxetable*}{cccccccc} 
  \tabletypesize{\footnotesize}
  \tablewidth{0pt}
  \tablecaption{Multiple image systems in MACS0647.\label{Table:MACS0647_1}}
\tablehead{
\colhead{ID} & \colhead{R.A.} & \colhead{Decl.} & \colhead{JWST} & \colhead{$z_{\mathrm{JWST}}$} & \colhead{HST} & \colhead{$z_{\mathrm{HST}}$} & \colhead{References\tablenotemark{a}}
}
\startdata
\hline
 1.1 & 101.966807 &  70.248308 & $\checkmark$ &   $ 2.134$ & $\checkmark$ & \nodata & 1,2\\ 
 1.2 & 101.955537 &  70.249205 & $\checkmark$ &            & $\checkmark$ & & \\ 
 1.3 & 101.960797 &  70.248533 & $\checkmark$ &            & $\checkmark$ & & \\ 
 1.4 & 101.966157 &  70.255822 & $\checkmark$ &            & $\checkmark$ & & \\ 
 1.5 & 101.952307 &  70.240001 & $\checkmark$ &            & $\checkmark$ & & \\ 
\hline
 2.1 & 102.001536 &  70.250343 & $\checkmark$ &   $ 4.750$ & $\checkmark$ & $ 4.7\pm0.5$ & 1,2,3\\ 
 2.2 & 102.001557 &  70.248537 & $\checkmark$ &            & $\checkmark$ & & \\ 
 2.3 & 101.994297 &  70.239374 & $\checkmark$ &            & $\checkmark$ & & \\ 
\hline
 3.1 & 101.974507 &  70.243384 & $\checkmark$ & $3.0\pm0.5$ & $\checkmark$ & \nodata & 1,2\\ 
 3.2 & 101.972537 &  70.242639 & $\checkmark$ &            & $\checkmark$ & & \\ 
\hline
 4.1 & 101.928176 &  70.249247 & $\checkmark$ & \nodata    & $\checkmark$ & \nodata & 1,2\\ 
 4.2 & 101.929006 &  70.245672 & $\checkmark$ &            & $\checkmark$ & & \\ 
 4.3 & 101.939047 &  70.257175 & $\checkmark$ &            & $\checkmark$ & & \\ 
\hline
 5.1 & 101.921077 &  70.251511 & $\checkmark$ & $6.5\pm0.5$ & $\checkmark$ & $6.5\pm0.5$ & 1,2\\ 
 5.2 & 101.921647 &  70.242900 & $\checkmark$ &            & $\checkmark$ & & \\ 
 5.3 & 101.934482 &  70.259419 & $\checkmark$ &            & & & \\ 
\hline
 6.1 & 101.971407 &  70.239714 & $\checkmark$ &   $10.170$ & $\checkmark$ & $10.8\pm0.5$ & 1,2,4,5\\ 
 6.2 & 101.982305 &  70.243261 & $\checkmark$ &            & $\checkmark$ & & \\ 
 6.3 & 101.981153 &  70.260572 & $\checkmark$ &            & $\checkmark$ & & \\ 
\hline
 7.1 & 101.962196 &  70.255523 & $\checkmark$ &   $ 2.135$ & $\checkmark$ & \nodata & 1,2\\ 
 7.2 & 101.952957 &  70.249959 & $\checkmark$ &            & $\checkmark$ & & \\ 
 7.3 & 101.948967 &  70.239780 & $\checkmark$ &            & $\checkmark$ & & \\ 
\hline
 8.1 & 101.952326 &  70.254267 & $\checkmark$ &   $ 2.170$ & $\checkmark$ & \nodata & 1,2\\ 
 8.2 & 101.947996 &  70.253630 & $\checkmark$ &            & $\checkmark$ & & \\ 
\hline
 9.1 & 101.932547 &  70.250090 & $\checkmark$ & $5.9\pm0.5$ & $\checkmark$ & $5.7\pm0.5$ & 1,2\\ 
 9.2 & 101.954486 &  70.260482 & $\checkmark$ &            & $\checkmark$ & & \\ 
 9.3 & 101.937547 &  70.239739 & $\checkmark$ &            & $\checkmark$ & & \\ 
\hline
10.1 & 101.919587 &  70.249039 & $\checkmark$ &   $ 7.460$ & $\checkmark$ & $7.0\pm0.5$ & 1,2,6\\ 
10.2 & 101.920586 &  70.244846 & $\checkmark$ &            & $\checkmark$ & & \\ 
10.3 & 101.936587 &  70.260541 & $\checkmark$ &            & & & \\ 
\hline
11.1 & 101.965127 &  70.246872 & $\checkmark$ & $2.2\pm0.5$ & $\checkmark$ & $2.4\pm0.5$ & 1,2\\ 
11.2 & 101.956047 &  70.242732 & $\checkmark$ &            & $\checkmark$ & & \\ 
11.3 & 101.967837 &  70.258380 & $\checkmark$ &            & $\checkmark$ & & \\ 
\hline
12.1 & 101.989842 &  70.244282 & $\checkmark$ & $2.8\pm0.5$ & & & 2\\ 
12.2 & 101.989085 &  70.243671 & $\checkmark$ &            & & & \\ 
\hline
13.1 & 101.989948 &  70.248422 & $\checkmark$ & $3.6\pm0.5$ & & & 2\\ 
13.2 & 101.988278 &  70.252389 & $\checkmark$ &            & & & \\ 
13.3 & 101.976408 &  70.238296 & $\checkmark$ &            & & & \\ 
\hline
14.1 & 102.002442 &  70.243880 & $\checkmark$ & $3.2\pm0.5$ & & & 2\\ 
14.2 & 102.002169 &  70.243670 & $\checkmark$ &            & & & \\ 
14.3 & 102.001525 &  70.242977 & $\checkmark$ &            & & & \\ 
\hline
15.1 & 101.999418 &  70.242435 & $\checkmark$ &   $ 3.230$ & & & 2,6\\ 
15.2 & 102.002495 &  70.247114 & $\checkmark$ &            & & & \\ 
15.3 & 102.002450 &  70.250209 & $\checkmark$ &            & & & \\ 
\hline
16.1 & 101.945672 &  70.248836 & $\checkmark$ & $6.7\pm1.0$ & & & 2\\ 
16.2 & 101.944823 &  70.248873 & $\checkmark$ &            & & & \\ 
16.3 & 101.968391 &  70.260522 & $\checkmark$ &            & & & \\ 
16.4 & 101.952456 &  70.239285 & $\checkmark$ &            & & & \\ 
\hline
17.1 & 101.971246 &  70.239714 & $\checkmark$ &   $10.170$ & & & 2,4,5\\ 
17.2 & 101.982239 &  70.243314 & $\checkmark$ &            & & & 2,4,5\\ 
17.3 & 101.981019 &  70.260602 & $\checkmark$ &            & & & \\ 
\hline
18.1 & 101.969857 &  70.239382 & $\checkmark$ &   $10.170$ & & & 2,4,5\\ 
18.2 & 101.982799 &  70.243861 & $\checkmark$ &            & & & 2,4,5\\ 
18.3 & 101.980493 &  70.260438 & $\checkmark$ &            & & & \\ 
\hline
19.1 & 101.990496 &  70.247681 & $\checkmark$ &   $ 3.600$ & & & 2\\ 
19.2 & 101.988028 &  70.253182 & $\checkmark$ &            & & & \\ 
19.3 & 101.975008 &  70.237833 & $\checkmark$ &            & & & \\ 
\hline
20.1 & 101.990804 &  70.247213 & $\checkmark$ & $3.6\pm0.5$ & & & 2\\ 
20.2 & 101.988125 &  70.253860 & $\checkmark$ &            & & & \\ 
20.3 & 101.974230 &  70.237479 & $\checkmark$ &            & & & %\\ 
%\hline
 \enddata
\tablenotetext{a}{1 -- \citet{2015ApJ...801...44Z}; 2 -- \citet{2023ApJ...944L...6M}; 3 -- \citet{2024MNRAS.527L...7F}; 4 -- \citet{2023ApJ...949L..34H}; 5 -- \citet{2024ApJ...973....8H}; 6 -- Worku et al. in prep.}
\end{deluxetable*}

\begin{deluxetable*}{cccccccc} 
  \tabletypesize{\footnotesize}
  \tablewidth{0pt}
  \tablecaption{Multiple image systems in MACS0647 ({\it continued}).\label{Table:MACS0647_2}}
\tablehead{
\colhead{ID} & \colhead{R.A.} & \colhead{Decl.} & \colhead{JWST} & \colhead{$z_{\mathrm{JWST}}$} & \colhead{HST} & \colhead{$z_{\mathrm{HST}}$} & \colhead{References\tablenotemark{a}}
}
\startdata
21.1 & 101.921269 &  70.246395 & $\checkmark$ & $1.5\pm0.5$ & & & 2 \\ 
21.2 & 101.921241 &  70.246093 & $\checkmark$ &            & & & \\ 
21.3 & 101.921555 &  70.245846 & $\checkmark$ &            & & & \\ 
\hline
22.1 & 101.961282 &  70.255530 & $\checkmark$ &  \nodata   & & & 2 \\ 
22.2 & 101.952220 &  70.250151 & $\checkmark$ &            & & & \\ 
22.3 & 101.948215 &  70.239818 & $\checkmark$ &            & & & \\ 
\hline
23.1 & 101.951590 &  70.250331 & $\checkmark$ &   $ 2.704$ & & & 2 \\ 
23.2 & 101.962494 &  70.256299 & $\checkmark$ &            & & & \\ 
23.3 & 101.947451 &  70.238350 & $\checkmark$ &            & & & \\ 
23.4 & 101.943942 &  70.251502 & $\checkmark$ &            & & & \\ 
\hline
24.1 & 101.950892 &  70.250447 & $\checkmark$ & $2.9\pm0.5$ & & & 2 \\ 
24.2 & 101.962165 &  70.256330 & $\checkmark$ &            & & & \\ 
24.3 & 101.947188 &  70.238392 & $\checkmark$ &            & & & \\ 
24.4 & 101.943728 &  70.251513 & $\checkmark$ &            & & & \\ 
\hline
25.1 & 101.960154 &  70.255867 & $\checkmark$ & $3.5\pm0.5$ & & & \\ 
25.2 & 101.944303 &  70.252327 & $\checkmark$ &            & & & \\ 
25.3 & 101.945858 &  70.237436 & $\checkmark$ &            & & & \\ 
25.4 & 101.951392 &  70.250894 & $\checkmark$ &            & & & \\ 
\hline
26.1 & 101.968233 &  70.250457 & $\checkmark$ & $3.4\pm0.5$ & & & \\ 
26.2 & 101.966753 &  70.253727 & $\checkmark$ &            & & & \\ 
26.3 & 101.949811 &  70.236126 & $\checkmark$ &            & & & \\ 
\hline
27.1 & 101.978425 &  70.253097 & $\checkmark$ &   $ 2.580$ & & & 2 \\ 
27.2 & 101.979869 &  70.249116 & $\checkmark$ &            & & & \\ 
27.3 & 101.961928 &  70.237782 & $\checkmark$ &            & & & \\ 
\hline
28.1 & 101.972548 &  70.238354 & $\checkmark$ & $5.1\pm0.5$ & & & \\ 
28.2 & 101.984543 &  70.256765 & $\checkmark$ &            & & & %\\ 
%\hline
 \enddata
\tablenotetext{a}{1 -- \citet{2015ApJ...801...44Z}; 2 -- \citet{2023ApJ...944L...6M}; 3 -- \citet{2024MNRAS.527L...7F}; 4 -- \citet{2023ApJ...949L..34H}; 5 -- \citet{2024ApJ...973....8H}; 6 -- Worku et al. in prep.}
\end{deluxetable*}

\section{Result of the Jackknife distributions with various set-ups}

\label{ap:jackknife_mcmc}

\begin{figure*}[p]
  \centering
  \includegraphics[width=18cm,clip]{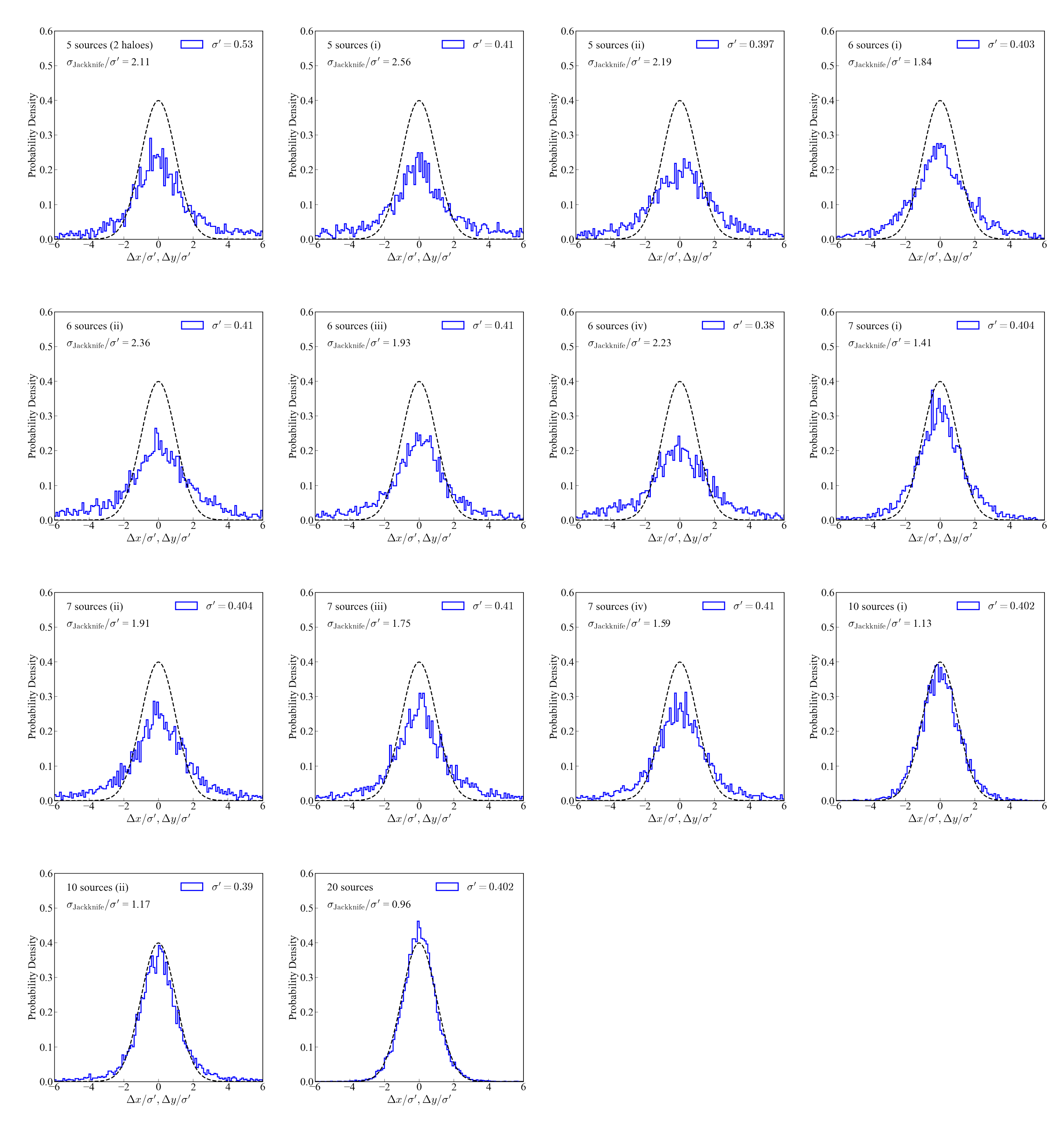}
  \caption{Result of the Jackknife distributions and $\sigma_{\rm{Jackknife}}/\sigma^{\prime}$ for incorrect models of all the models in Table~\ref{Table:Setting_input_fitting}. The Roman numbers distinguish different realizations of source positions.
  }
  \label{fig:jackknife_all}
\end{figure*}

Figure~\ref{fig:jackknife_all} shows the Jackknife distributions and the ratio $\sigma_{\rm{Jackknife}}/\sigma^{\prime}$ for the incorrect models of all set-ups listed in Table~\ref{Table:Setting_input_fitting}.

\end{appendix}

\end{document}